\documentclass[12pt]{article}
\newcommand\be{\begin{equation}}
\newcommand\ee{\end{equation}}
\newcommand\ba{\begin{eqnarray}}
\newcommand\ea{\end{eqnarray}}
\newcommand\eq{\begin{equation}}           
\newcommand\en{\end{equation}}

\newcommand\sn{ \widetilde{N}_1}
\newcommand\snc{ \widetilde{N}_{c}}
\newcommand\snstar{ \widetilde{N}_{*}}
\newcommand\snd{ \widetilde{N}_{d}}
\newcommand\phim{\phi_-}
\newcommand\phip{\phi_+}
\newcommand\GeV{\mbox{GeV}}
\input epsf
\usepackage{amssymb}
\begin{document}
\title{
{\hfill  \small  FERMILAB-PUB-05-544-A, RESCEU-41/05}
\\
~
\\
D-term Inflation and Leptogenesis \\ by Right-handed Sneutrino}

\author{
Kenji Kadota$^1$ and J. Yokoyama$^2$\\
$^1$ {\em Particle Astrophysics Center, Fermilab, Batavia, IL 60510,
USA} \\
$^2$ {\em Research Center for the Early Universe (RESCEU),} \\
{\em Graduate School of Science, The University of Tokyo, Tokyo, 113-0033, Japan}
}
\maketitle   
\begin{abstract}
We discuss a D-term inflation scenario where a right-handed sneutrino can be an inflaton field leading to a viable inflation and leptogenesis, with a minimal form of K\"ahler potential. The decay of an inflaton sneutrino can non-thermally create large enough lepton asymmetry. Its entropy production is also big enough to ameliorate the gravitino problem caused by too high a reheating temperature from the decay of a symmetry breaking field.
\end{abstract}

\setcounter{footnote}{0} 
\setcounter{page}{1}
\setcounter{section}{0} \setcounter{subsection}{0}
\setcounter{subsubsection}{0}

\section{Introduction}
Supersymmetry generically predicts the existence of fundamental scalar 
fields and those scalar fields can potentially play significant roles in
the early universe. Inflation is one of those possible early universe
phenomena \cite{gliner, Sato:1980yn}, and, in addition to the motivation for introducing scalar fields, supersymmetry is crucial to maintain the flatness of an inflaton potential by cancellation of radiative corrections from bosons and fermions \cite{tony}. Even with such features of supersymmetry, however, keeping the flatness of potential is highly non-trivial. In particular, in local supersymmerty (a.k.a. supergravity (SUGRA)), the scalar potential becomes, in terms of K\"ahler potential $K$ and super potential $W$,
\ba
V=e^K\left(W_i(K^{-1})^i_jW^{j} -3|W|^2\right)+\mbox{D-terms}
\ea
where $W_i=\partial_{\phi_i}W+W\partial _{\phi_i}K$, and $(K^{-1})^i_j$
is the inverse of the matrix
$\partial^2_{\phi_i,\phi^{*j}}K$. For F-term inflation \cite{f1,f2} where F-term dominates during inflation, the second derivative of the potential becomes (prime denotes the derivative with respect to an argument, in this case, an inflaton field)
\ba
V''=K''V+\mbox{other terms}
\ea
and, after canonical normalization $K''=1$,
\ba
\frac{V''}{V}=1+\mbox{other terms}
\ea
So one of the slow-roll conditions $|V''/V|\ll 1$ is generically violated unless other additional terms of order unity turn out to be canceled out or we choose a special form of K\"ahler potential\footnote{Note, however, that $|V''/V|\ll 1$ is not a necessary condition for producing the flat spectrum. See, for example, Ref. \cite{ks} for a concrete supergravity inflation model with $|V''/V|\gtrsim 1$.}. On the other hand, D-term inflation \cite{hal,dvali,casa,ewan}, where D-term is responsible for the driving energy of inflation, is free from the above mentioned problem because the gauge invariance prohibits the coupling of K\"ahler potential to D-terms. D-term inflationary scenario is therefore an interesting possibility to investigate in the framework of supergravity. Moreover, from the viewpoint of particle phenomenology, it is intriguing to identify an inflaton field to be one of the fields which show up in the supersymmetric Lagrangian and are related to other supersymmetric standard model fields. We, in this letter, present our discussion of D-term inflation in view of the application to the right-handed sneutrino inflation. Among the fields in the minimal extension of supersymmetric Lagrangian with the see-saw mechanism \cite{seesaw} motivated from neutrino physics, right-handed (s)neutrinos are special for the hierarchically large masses and they can play non-trivial roles in the early universe. In particular, we consider a scenario, where, with a minimal form of K\"ahler potential, a right-handed sneutrino can play a role of an inflaton field and its decay after inflation can create large lepton asymmetry via non-thermal leptogenesis.  

Let us briefly mention the features of the D-term sneutrino inflation model we will discuss in this paper compared with other sneutrino inflation scenarios. Simple chaotic sneutrino inflation models \cite{yokoyama,ellis} are capable of producing the desired flat CMB spectrum with the mass of right-handed sneutrino $M_N\sim 10^{13} \GeV$ if one uses a non-minimal K\"ahler potential. Note that, in such a heavy sneutrino mass, merely reducing the Yukawa coupling to lower the reheating temperature may not work because the gravitational coupling can become more important to lead to a high reheating temperature $T \sim 0.1 \sqrt{M_N^3}\sim 10^{9}$ GeV for $M_N \sim 10^{13} \GeV$ unless the gravitational couplings are further suppressed. Such a model potentially suffers from the over-production of gravitinos unless there exists late-time entropy production to dilute the unwanted relics. Our scenario works for a relatively small right-handed neutrino mass, say $M_N \sim 10^{10}\GeV$, and the reheating temperature from sneutrino decay can be relatively low. Another possibility, besides a chaotic inflation model, would be a hybrid inflation model \cite{lindehy}. A hybrid inflation model by a sneutrino inflaton field was proposed in the framework of F-term inflation \cite{shafi1}, but it requires tuning or/and specific form of K\"ahler potential to maintain the flatness of the potential, a typical problem of F-term inflation as discussed above (so-called $\eta$ problem). D-term inflation, however, typically suffers from a high reheating temperature due to the decay of a symmetry breaking field, but it is ameliorated by the entropy production from the decay of inflaton sneutrino as we shall discuss.

The paper is structured as follows. In Sec. \ref{before}, we present our model and discuss the dynamics of inflation. We point out that, in D-term inflation models, the SUGRA corrections for a relatively large inflaton amplitude completely change the analytical expression for the cosmic perturbations compared with the cases of a relatively small inflaton amplitude. In Sec. \ref{after}, we describe the post-inflation dynamics and examine the lepton asymmetry production from the decay of inflaton sneutrino and argue how the reheating temperature constraint is relaxed to avoid the gravitino problem, followed by Conclusion in Sec. \ref{dis}.

\section{D-term Inflation Dynamics: Cosmic Perturbations and Strings}
\label{before}

We consider the supersymmetric model with $U(1)$ gauge group under which the chiral superfields $\phi_{\pm}$ have the charges $q_{\pm}=\pm 1$, and the superpotential relevant for right-handed neutrinos has the form 
\ba
W\supset \frac\lambda M_* N_i^cN_i^c \phi_- \phi_+ + \frac12 M_iN_i^cN_i^c +h_{i,\alpha}N_i^cL_{\alpha}H_u
\ea
$N_i^c,L_{\alpha}$ an $H_u$ are the superfields containing the
right-handed (s)neutrino, the left-handed (s)lepton doublets and
Higgs/Higgsino fields respectively.  $M_*$ can be the Planck scale or, if
this non-renormalizable term comes from integrating out some heavy
degree of freedom, the mass of a heavy field. For definiteness, we
present our discussion with $M_*=M_{G}$ and we use the Planck units
treating the reduced Planck scale $M_{G}=2.4\times 10^{18}\GeV$ to be
unity. $h_{i,\alpha}$ is Yukawa coupling and $i$ runs for $1\sim 3$ and
$\alpha=e,\nu,\tau$. Right-handed neutrinos have odd R-parity which
prohibits the cubic term $(N^c_i)^3$ (we assume R-parity is
exact)\footnote{The standard D-term inflation models have the form of
$W\supset N^c \phi_- \phi_+$. This form however leads to too big a Dirac
mass to explain the expected small neutrino mass by the standard simple
see-saw mechanism. The absence of such a term can be justified by, for example, assigning the same R-parity to $\phi_{\pm}$.}. One can always choose a basis for $N_i^c$ so that
their mass matrix is diagonal, and we take $N_i^c$ to be Majorana mass
eigenstate fields with real mass $M_i$. We assume, without loss of
generality, the inflaton sneutrino is the lightest heavy sneutrino
$M_1\ll M_2,M_3$ and we are interested in the lower range of the
preferred values of heavy neutrino masses $M_i=10^{10}\sim
10^{15}\GeV$ \footnote{Our inflation/leptogenesis scenario can work for an even smaller sneutrino mass as well.}. With an appropriate R transformation, the scalar
components of $N_1^c$ and $\phi_{\pm}$ can be made to be real and 
we consider, for simplicity, the real components.  Hereafter
$\sn$ and ${\phi}_{\pm}$ denote the real parts of the complex 
scalar fields $\widetilde{N}^c$ and $\phi_\pm$ multiplied by
$\sqrt{2}$ so that they are canonically normalized.

D-term contribution to the scalar potential with non-vanishing Fayet-Illiopoulos (FI) term $\xi$($>0$. We can just change the roles of $\phi_{\pm}$ for $\xi<0$) reads
\ba
V_D=\frac12g^2(\xi-|\phi_-|^2+|\phi_+|^2)^2
\ea
The large amplitude of inflaton sneutrino gives a large effective mass to $L$ and $H_u$ and they stay at the origin and do not affect the inflation dynamics. $\phip$ has a positive mass during and after inflation and it stays at the origin all the time. For the discussion of inflationary dynamics, therefore, we discuss the evolution of $\sn$ and $\phim$.

During inflation, the mass of $\phi_-$ is
\ba
m^2_{\phi_-}=\frac{\lambda^2 \sn^4}{4}+\frac{M_1^2\sn^4}{16}-g^2\xi
\ea
$\lambda\gg M_1$, because we are interested in the parameter 
regime of $M\sim 10^{10}\GeV$, so that $\phi_-$ stays at the origin for 
$\sn>\snc \equiv (4g^2\xi/\lambda^2)^{1/4}$ during inflation and 
destabilizes for $\sn< \snc $ to reach its minimum at
$\phim=\sqrt{2\xi}$ 
after inflation.

The scalar potential for $\sn$ during inflation with the other scalar 
fields at the origin can be given, for the first order approximation, 
as a sum of SUGRA potential and radiative corrections.
 SUGRA potential is 
\ba
V_{SUGRA}(\sn)=e^{\sn^2/2}\left(\frac{1}{2}M_1^2\sn^2
+\frac{1}{16}M_1^2\sn^4+\frac{1}{32}M_1^2\sn^6\right)+\frac12g^2\xi^2
\ea
 We consider the minimal K\"ahler potential because one of our purposes of this section is to show that a successful inflationary scenario can be realized even with a simple minimal form of K\"ahler potential. The effects of non-minimal K\"ahler potential are discussed, for example, in Ref. \cite{yokoyama} for the sneutrino inflation and Ref. \cite{sy} for D-term inflation.

Radiative corrections come from the mass splitting in the boson and fermion partners of $\phi_\pm$ due to SUSY breaking, and it is \cite{coleman}
\ba
\label{rad}
V_{rad}(\sn)\simeq \frac{g^4\xi^2}{16\pi^2} \ln \left[e^{\sn^2/2}\frac{\lambda^2\sn^4}{4\Lambda^2}\right]
\ea
where $\Lambda$ denotes the renormalization scale, and, in the above expression, we assumed $\lambda\gg M_1$ and $\sn^4\lambda^2 \gg g^2 \xi$ which stabilizes $\phim$ at the origin.

We are interested in the parameter range $M_1^2\ll g^2\xi^2$, so
 that the sneutrino potential becomes flat enough for the flat cosmic perturbation spectrum. For this case, then the F-I term dominates the energy of the universe inducing the hybrid inflation \cite{lindehy}, and the derivatives of $V(\sn)=V_{SUGRA}(\sn)+V_{rad}(\sn)$ can be approximated by those of $V_{rad}(\sn)$
\ba
\label{slope}
V'(\sn)&=&\frac{g^4\xi^2}{16\pi^2} \left(\sn +\frac{4}{\sn}\right)\\
V''(\sn)&=&\frac{g^4\xi^2}{16\pi^2} \left(1 -\frac{4}{\sn^2}\right)
\ea
where the prime denotes the derivatives with respect to the inflaton sneutrino.
 Inflation ends when $\sn= \snc
     $ or when the slow-roll conditions are violated, whichever comes first.

The parameters can be constrained by the amplitude of the comoving curvature perturbation ${\cal R}_c$ 
produced when the cosmologically interesting scales leave the horizon. Letting this epoch occur when $\sn=\widetilde{N}_{*}$ which is the value of $\sn$ at ${\cal N}$ e-folds before the end of inflation,
\ba
\label{curv2}
{\cal R}_c=\frac{H^2}{2\pi |\dot{\sn}|}=\frac{3H^3}{2\pi|V'|}=\frac{2\sqrt2}{\sqrt3}\frac{\xi \pi}{g} \left(\widetilde{N}_{*}+\frac{4}{\widetilde{N}_{*}}\right)^{-1}
\ea
where we used the standard slow-roll approximations. The observation requires ${\cal R}_c=4.7\times 10^{-5}$.
This has a potential conflict with the constraints coming from the cosmic strings which form due to the the tachyonic nature of $\phim$ at the end of inflation with the string tension $\sim 2 \pi \xi$. For the cosmic strings not to make too significant contributions to the observable cosmic perturbation spectrum, one requires \cite{string,rocher,endo}
\ba
\label{cosmic}
 \sqrt{\xi}\lesssim 4.6 \times 10^{15} \GeV
\ea

Let us point out that this tension between the cosmic perturbation constraints (which tend to require a
relatively large value of $\xi$) and the cosmic string constraints (which
tend to require a relatively small value of $\xi$) is relaxed for our model of the form $W\supset N^c N^c \phi_- \phi_+$ compared with the standard D-term inflation of the form $W\supset N^c \phi_- \phi_+$ due to the different dependence of ${\cal R}_c$ on $\xi$.

\subsection{General Solution}
Using the standard slow-roll approximation, the equation of motion in terms of the time variable $\ln a$ ($a$ is a scale factor) becomes
\ba
 V(\sn) \frac{d\sn}{d\ln a}+V'(\sn)=0
\ea
Note $d\ln a=-d{\cal N}$. Substituting Eq (\ref{slope}) and $V=g^2\xi^2/2$, the solution for the equation of motion for $\widetilde{N}_{*}$ becomes
\ba
\label{general}
\frac{4+\widetilde{N}_{*}^2}{4+\widetilde{N}_{f}^2}=\exp\left({\frac{g^2}{4\pi^2}{\cal N}}\right)
\ea
where $\widetilde{N}_{f}$ is the amplitude of $\sn$ when inflation terminates.

As a concrete example, let's consider the case of
${\cal N}=55, \sqrt{\xi}=4.6\times 10^{15}\GeV,g=0.1$. COBE/WMAP
normalization ${\cal R}_c\sim 4.6\times 10^{-5}$ using Eq (\ref{curv2})
leads to $\lambda= 10^{-4}$. Inflation terminates when $\sn=\snc=1.94$
and the cosmologically interesting scales leave the horizon when
$\snstar=1.97$, leading to the slow-roll parameters $\epsilon\equiv 1/2
(V'/V)^2\sim 1.2\times 10^{-7},\eta\equiv V''/V \sim -4\times 10^{-6}$
for the flat spectrum. For the parameters in this example, in order
 that F-I term dominates the energy density of the Universe, we need $M^2\snstar^2\ll g^2\xi^2$ or $M_1\ll 4.3\times 10^{11}\GeV$.

The expressions for the curvature perturbations can be simplified for the small or large inflaton amplitudes, and we discuss below those approximations for the illustration purposes.
\subsection{Special Case I: Small Inflaton Amplitude}
\label{smalln}
We now discuss the case when the sneutrino amplitude is comparable or less than Planck scale (more precisely, $\sn \lesssim 2$) while the cosmologically interesting scales leave the horizon, then the solution of equation of motion (Eq (\ref{general})) can be approximated as 
\ba
\label{sol1}
{\widetilde{N}_{*}^2}-{\widetilde{N}_{f}^2}=\frac{g^2}{\pi^2}{\cal N}
\ea
The comoving curvature perturbation ${\cal R}_c$ produced when $\sn=\widetilde{N}_{*}$ can be approximated as, from Eq (\ref{curv2}),
\ba
\label{sol11}
{\cal R}_c=\frac{\pi \xi \widetilde{N}_{*}}{\sqrt6 g}
\ea
$\phi_-$ destabilizes ending inflation when $\widetilde{N}_{f} =\widetilde{N}_{c}$ which occurs before the slow-roll condition violates 
for the parameter range of our interest $g\lambda/\sqrt{\xi}\lesssim 4\pi^2$, and ${\cal R}_c$ therefore can be approximated as, substituting $\widetilde{N}_{f}= \snc\equiv (4g^2\xi/\lambda^2)^{1/4}$ to Eqs (\ref{sol1},\ref{sol11}),
\ba
{\cal R}_c=
\frac{\xi\sqrt{{\cal N}}}{\sqrt6}\sqrt{1+\frac{2\pi^2 \sqrt{\xi}}{g\lambda {\cal N}}}
\ea

Hence
\begin{eqnarray}
{\cal R}_c
\simeq 
\left\{
\begin{array}{ll}
\frac{\xi \sqrt{{\cal N}}} {\sqrt{6}} & ~~~\mbox{for  } \lambda\gg \lambda_c\equiv \frac{2\pi^2 \sqrt{\xi}}{g{\cal N}}
\\
\frac{\pi}{\sqrt3}\frac{\xi^{5/4}}{\sqrt{g\lambda}} &~~~\mbox{for  } \lambda\ll\lambda_c 
\end{array}
\right .
\label{pert2}
\end{eqnarray}
For $\lambda\gg \lambda_c$, the curvature perturbations can be approximated as 
\ba
{\cal R}_c=4.7\times 10^{-5}\left(\frac{\sqrt{\xi}}{9.5\times 10^{15}\GeV}\right)^2 \ \sqrt{\left(\frac{{\cal N}}{55}\right)}
\ea
This is incompatible with the cosmic string constraint given by Eq (\ref{cosmic}).
So the regime of $\lambda\ll \lambda_c$ would be of our interest where 
\ba
\label{curv1}
{\cal R}_c=\frac{\pi}{\sqrt3}\frac{\xi^{5/4}}{\sqrt{g\lambda}}
\ea

At first sight, Eq (\ref{curv1}) seems to indicate that we can simultaneously satisfy both string and perturbation amplitude constraints by lowering the value of $\lambda$ for a given $\xi$ as often stated in literature. Lowering $\lambda$ for a given  $\xi$  however makes $\snc= (4g^2\xi/\lambda^2)^{1/4}$ larger and the above expression cannot be applied if $\sn\gtrsim 2$. We consider such a large amplitude case in the following subsection.

\subsection{Special Case II: Large Inflaton Amplitude}

We next discuss the case when the sneutrino field has a large amplitude beyond Planck scale $\sn\gtrsim 2$, where the contribution from K\"ahler potential becomes non-negligible. For a large inflaton amplitude, from Eq (\ref{general}), the solution for the equation motion can be approximated as
\ba
\frac{\widetilde{N}_{*}}{\widetilde{N}_{f}}=\exp\left(\frac{g^2}{8\pi^2}{\cal N}\right) 
\ea
${\cal R}_c$ then becomes, using Eq (\ref{curv2}),
\ba
{\cal R}_c=\frac{2\sqrt2}{\sqrt3} \pi\frac{\xi}{g\widetilde{N}_{*}}=\frac{2\sqrt2}{\sqrt3} \pi  \frac{\xi}{g \widetilde{N}_{f}}  \exp\left( -\frac{g^2}{8\pi^2}   {\cal N} \right)   
\ea
$\phi_-$ destabilizes ending inflation for $\widetilde{N}_{f} =\widetilde{N}_{c}$ which occurs before the slow-roll condition is violated 
for the parameter range of our interest $g^5 \sqrt{\xi} / \lambda \lesssim  64\pi^4$, and ${\cal R}_c$ therefore becomes
\ba
\label{cobe}
{\cal R}_c=\frac{2\pi}{\sqrt3}\frac{\xi^{3/4}\sqrt{\lambda}}{g^{3/2}} \exp\left( -\frac{g^2}{8\pi^2}   {\cal N} \right)       
\ea
As we mentioned at the end of Sec. \ref{smalln}, lowering $\lambda$ does not help to make $\xi$ small to circumvent the cosmic string constraint. Reducing the value of $\lambda$ increases $\snc$ and consequently also the value of $\sn$ relevant for inflation dynamics of our interest, making the K\"ahler potential contribution non-negligible. The same consideration applies to the standard simple D-term inflation scenarios where the superpotential has a form $W=\lambda S\phi_-\phi_+$ instead of $W=\lambda SS\phi_-\phi_+$ as discussed here \cite{rocher}.

\section{Dynamics after Inflation: Leptogenesis and Gravitino Problem}
\label{after}
After inflation, the potential has zero minimum for $\sn=0,\phi_+=0$ and $\phim=\sqrt{2 \xi}$. The field oscillations around the minimum are characterized by, for the oscillation along $\sn$ direction, the mass of $\sn$ 
\ba
m_{\sn}^2=\frac{\partial^2 V}{\partial \sn^2}(\sn=0,\phip=0,\phim=\sqrt{2\xi})=M_1^2
\ea
and for $\phim$ direction 
\ba
m_{\phim}^2=\frac{\partial^2 V}{\partial \phim^2}(\sn=0,\phip=0,\phim=\sqrt{2\xi})=
2\xi g^2
\ea

The mass scale $M_1$ is related to the light neutrino masses 
via the see-saw mechanism. The term in the superpotential 
$N^cN^c \phi_- \phi_+$ gives neither 
Dirac nor Majorana masses to neutrinos, and we have the 
standard see-saw mechanism \cite{fuku} 
\ba
(m_{\nu})_{\alpha \beta}=-\sum_i h_{i\alpha}h_{i\beta}\frac{\langle H_u \rangle^2}{M_i}
\ea

The lepton asymmetry arises from the $\sn$ decay into lepton
 $L\widetilde{H} _u$ and that into anti-lepton $\widetilde{L}^*{H} _u^*$
 \cite{yokoyama,fuku,cp,olive1}
\ba
\label{epsi}
\epsilon_1&\equiv& \frac{\Gamma(\sn\rightarrow L +\widetilde{H}_u)-\Gamma(\sn \rightarrow \widetilde{L}^*+H_u^*)}
{\Gamma(\sn\rightarrow L +\widetilde{H}_u)+\Gamma(\sn \rightarrow \widetilde{L}^*+H_u^*)} \nonumber \\
&\simeq& -\frac{3}{8\pi} \frac{1}{(hh^{\dagger})_{11}} \sum_{i=2,3}{\mbox{Im}}[(hh^{\dagger})^2_{1i}]\frac{M_1}{M_i}
\ea
Equation (\ref{epsi}) can be written as, 
in terms of the mass of the heaviest left-handed neutrino $m_{\nu_3}$,
\ba
\label{bary}
\epsilon_1&=&\frac{3}{8\pi}\frac{M_1}{\langle H_u \rangle^2}m_{\nu _3} \delta_{eff}\\
&\approx&2\times
10^{-6}\left(\frac{M_1}{10^{10}\GeV}\right)\left(\frac{m_{\nu
_3}}{0.05{\rm eV}}\right)\left(\frac{174\GeV}{\langle H_u \rangle}\right)^2 \delta_{eff}
\ea
where, for the quantitative estimation of baryon asymmetry, we used the
representative values of $\langle H_u \rangle \approx 174\GeV$ and
$m_{\nu_3}\approx 0.05$ eV
 \cite{sk}, and the effective CP violating phase is
\ba
\delta_{eff} =\frac{{\mbox{Im}}[h(m^*_{\nu})h^T]_{11}}{m_{\nu_3}(hh^{\dagger})_{11}}
\ea
We are discussing the inflation by the lightest heavy singlet sneutrino and the lepton asymmetry produced by the heavier sneutrinos are washed out due to the lepton number violating processes of $\sn$ in the parameter range of our interest (we will later discuss the reheating by $\phim$ decay, but the lepton asymmetry produced there will also be washed out.). We also assume $|M_1-M_i|\ll|\Gamma_1-\Gamma_i|$ for $i=2,3$ ($\Gamma$ is the decay rate) for the validity of the perturbative calculation in estimating the baryon asymmetry\footnote{We shall not discuss other special cases such as the case of the almost degenerate neutrino masses \cite{res} for which the baryon asymmetry can be bigger than our estimates and hence the constraints on the parameters in our scenario can be relaxed.}.

After the inflation, $\sn$ decays when its decay rate 
\ba
\label{decayrate}
\Gamma_{\sn}=\frac{M_1}{8\pi}\sum_{\alpha}|h_{1\alpha}|^2
\ea
becomes of the same order as the expansion rate
\ba
\label{density}
H\approx \sqrt{\frac{\pi^2}{90}g_*T_1^4}
\ea
where the effective number of degrees of freedom $g_*\sim 230$ when $\sn$ decays at temperature $T_1\gg 1$TeV. $\Gamma_{\sn}=H$ gives
\ba
\label{nreh}
T_1&=&\left(\frac{90}{\pi^2 g_*}\right)^{1/4} \sqrt{\Gamma_{\sn}} \\
&=&1.4\times 10^{10}\GeV\sqrt{\left(\frac{M_1}{10^{10}\GeV}\right)}
\sqrt{\left(\frac{\sum_{\alpha}|h_{1\alpha}|^2}{(10^{-3})^2}\right)}
\ea 
where we assumed the Yukawa coupling is the dominant decay channel for $\sn$.
The reheating temperature is constrained from the gravitino production\footnote{We do not consider the non-thermal production of gravitinos because of its strong model dependence, but it most likely would not lead to the tighter constrains than that of the thermal gravitino production for a wide class of inflation models \cite{marco}.} whose abundance is approximately proportional to the reheating temperature \cite{moroi}
\ba
\label{gra}
\frac{n_{3/2}}{s}\approx 1.5\times 10^{-12}\left(\frac{T_R}{10^{10}\GeV}\right)
\ea
because they can jeopardize the successful nucleosynthesis if they decay during/after nucleosynthesis \cite{oldgrav1,grav2}.
The recent analysis including the hadronic decay indicates one needs
$T_R<10^{6\sim7}\GeV$ for the $10^{2\sim3}\GeV$ mass gravitinos
\cite{moroi}, if one assumes no dilution of gravitinos until the
nucleosynthesis. Note, for such a 'safely' low reheating temperature,
$M_1>T_1$ so that the decay of $\sn$ occurs out of equilibrium. The
above simple estimate suggests the preferable range of Yukawa couplings
for $\sn$ is $\lesssim 10^{-6}$ for $M_1\sim 10^{10}$GeV.  We will also discuss the case of 
the existence of late-time entropy productions at the end of section,
as they dilute gravitinos and baryon asymmetry of the universe 
where our scenario does not necessarily require such a small Yukawa
coupling.

Using the entropy density
\ba
s=\frac{2\pi^2}{45}g_*T_1^3
\ea
the lepton number to entropy ratio becomes
\ba
\frac{n_L}{s}=\frac34\epsilon_1\frac{T_1}{M_1}
\ea
The sphaleron effects then convert it to baryon asymmetry
\ba
\frac{n_B}{s}\sim c_{sph} \frac{n_L}{s}
\ea
with $c_{sph}\sim -8/23$ \cite{mike}. The baryon asymmetry therefore becomes
\ba
\label{baryo}
 \frac{n_B}{s} \sim 1.5 \times 10^{-10}
 \left(\frac{T_1}{10^6\GeV}\right) \left(\frac{m_{\nu_3}}{0.05{\rm eV}}\right) \left(\frac{174\GeV}{\langle H_u \rangle}\right)^2 c_{sph} \delta_{eff}
\ea
Unless $\delta_{eff}$ is very small, we can expect the right order of baryon asymmetry even with a relatively low reheating temperature.

We also need to consider the possible effects from the decay of $\phim$. For a pessimistic estimation, where we assume the efficient thermalization of $\phim$ decays\footnote{We also ignore possible suppressions of the decay such as the effect of large vev of $\phim$ giving large masses to other fields coupling to $\phim$, which turns out not be so strong suppressions because those effective masses are comparable to $m_{\phim}$.}, the decay rate of $\phim$ is
\ba
\Gamma_{\phim}\sim \frac{g^2}{8\pi} m_{\phim}=\frac {g^3}{8\pi} \sqrt{2\xi}
\ea
which leads to the reheating of the universe by $\phim$ decay 
\ba
T_{\phim}=\left(\frac{90}{\pi^2 g_*}\right)^{1/4} \sqrt{\Gamma_{\phim}}=1.7 \times 10^{15}\GeV\sqrt{\left(\frac{g}{0.1}\right)^3\left(\frac{\sqrt{\xi}}{4.6\times 10^{15} \GeV}\right)}
\ea
So the decay of $\phim$ can potentially lead to a rather high reheating
temperature to overproduce the gravitinos upsetting the nucleosynthesis,
which is a common problem in simple D-term inflation
models\footnote{See, however, Ref. \cite{kolda} for a possibility to
reduce the reheating temperature greatly by considering the suppression
of the decay of $\phim$ to MSSM fields.}. But noting $\sn$ decays much
later than $\phim$ decays, this apparent problems can be ameliorated by considering the gravitino dilution thanks to the entropy production from inflaton sneutrino decay \cite{kolda, ndominated}. Depending on the parameter ranges of our model, we can consider the following cases.

1) If the decay of $\phim$ is completed by the time when $\sn$ starts oscillation (so that the Universe is radiation dominated by the decay products of $\phim$ when $\sn$
starts oscillation), the dilution factor due to $\sn$ decay is
\ba
\Delta_1 \sim 10^8 \sqrt{\left(\frac{M_1}{10^{10}\GeV}\right)}\left(\frac{10^6 \GeV}{T_1}\right)\snd^2
\ea 
where $\snd$ is the initial amplitude of $\sn$ oscillation.

2) We can also consider the case when $\phim$ is still oscillating when $\sn$ starts oscillation\footnote{If one needs to follow the dynamics of the oscillation when the amplitudes of the oscillations for both fields happen to be comparable, it would be more appropriate to follow the dynamics of a field consisting of the linear combination of these two fields.}. Even in this case, $\phim$ decays much before the decay of $\sn$ for the parameter range of our interest, and the dilution factor becomes
\ba
\Delta_1\sim 10^8
\left(\frac{T_{\phim}}{10^{14}\GeV}\right)
\left(\frac{10^6 \GeV}{T_1}\right)
\snd^2
\ea
In both cases above, we assumed $\sn$ oscillation (whose energy is
red-shifting as the matter) dominates the energy of the universe when
$\sn$ decays, which is reasonable because $\Gamma_{\phim}\gg
\Gamma_{\sn}$ and the decay products from $\phim$ decay are red-shifting
away as 
radiation\footnote{We shall not discuss the possible pre-heating effects which can convert the energy of a classical field to the corresponding particles efficiently \cite{linde1}. The reheating temperature should be calculated by thermal processes, but non-thermal pre-heating effects still may affect the subsequent reheating temperature estimations. See also Ref. \cite{turb} for the estimation of the reheating temperature taking account of the detailed thermalization processes.}.  
If there is an entropy dilution as in these cases, the reheating temperature constraints from the gravitino problem should be applied to $T_{\phim}/\Delta_1$ rather than to $T_{\phim}$ (see Eq (\ref{gra})). So, for instance, if we have $\Delta_1\sim 10^{8}$, the reheating temperature as high as $T_{\phim} \sim 10^{14\sim 15}\GeV$ can still be allowed from the $\phim$ decay even for the typical gravitino mass range $10^{2\sim 3}\GeV$, and our model can give a viable inflationary scenario. 

Of course, the reheating temperature constraint is relaxed if we make different assumptions on the gravitino mass ranges (e.g. stable or very heavy gravitinos). It can also be relaxed if there exist additional entropy productions after $\sn$ decay. Indeed, in existence of late-time entropy production, a very small Yukawa coupling is not necessarily required. The dilution of gravitinos and baryon asymmetry due to the late entropy production is not so unnatural considering that there are many flat directions in the supersymmetric field space. Those scalar fields will be displaced from their minimum along the flat directions due to quantum fluctuations during inflation, and their subsequent oscillations around the minima and their consequent decays may produce the non-negligible amount of entropy. 
For example, if the late-time entropy productions give the dilution
factor of $\Delta\sim 10^{3\sim 4}$, $T_1\sim 10^{9\sim 10} \GeV$ can be
possible\footnote{We still need $M_1\gtrsim 0.1T_1$ to ensure the out-of-equilibrium decay of $\sn$ for the non-thermal leptogenesis.}. We also need to check the the baryon asymmetry which is also diluted by such a late-time entropy production. This can be easily checked from Eq (\ref{baryo})
\ba
 \frac{n_B}{s} \sim 1.5 \times 10^{-10} \left(\frac{T_1}{10^{10}
 \GeV}\right) \left(\frac{m_{\nu_3}}{0.05{\rm eV}}\right) \left(\frac{10^4}{\Delta}\right) \left(\frac{174\GeV}{\langle H_u \rangle}\right)^2 c_{sph} \delta_{eff}
\ea
which shows that our model is capable of creating a large enough baryon asymmetry of the universe even with late-time entropy production, relaxing the tuning of Yukawa couplings.

\section{Conclusion}
\label{dis}
We presented a sneutrino inflation scenario where the D-term dominates the energy density of the universe during inflation. We showed that the addition of a simple term $W\supset N_i^cN_i^c\phi_-\phi_+$ to the standard Majorana and Dirac mass terms $W\supset M_iN_i^cN_i^c+N_i^cL_{\alpha}H_u$ can realize a viable hybrid inflation and leptogenesis in supergravity, while keeping the standard form of see-saw formula to explain the small left-handed neutrino masses. The realization of inflation using a sneutrino field which shows up naturally in the minimal extension of supersymmetric standard model would be attractive from the viewpoint of particle phenomenology. In particular, leptogenesis is well motivated from neutrino physics as well. One of the advantages of inflation induced by a sneutrino is that the reheating temperature which is a crucial quantity parameterizing the amount of produced lepton asymmetry in leptogenesis scenarios is related to the properties of a sneutrino. So observing the current baryon asymmetry of the universe can be directly related to the sneutrino properties in contrast to other leptogenesis scenarios where inflation reheating temperature obscures the direct relation between the observed baryon asymmetry and neutrino properties.

\subsection*{Acknowledgments}  
We thank Mu-Chun Chen, Dan Chung, Andre de Gouvea, Rocky Kolb and Albert Stebbins for the useful discussions, and KK thanks RESCEU at University of Tokyo for the hospitality where this work was initiated. KK was supported by DOE and by NASA grant NAG5-10842. JY was supported by the JSPS Grant-in-Aid for Scientific Research No. 16340076.



\begin{thebibliography}{99}
\bibitem{gliner}
E. Gliner,
\textit{Algebraic properties of the energy-momentum tensor and vacuum-like states
of matter},
Sov. Phys. Zh. Eksp. Teor. Fiz. {\bf 49} (1965) 542 [JETP {\bf 22} (1966) 378];
\textit{The vacuum-like state of a medium and Friedmann cosmology},
Sov. Phys. Dokl. {\bf 15} (1970) 559;
E. Gliner and I. Dymnikova,
\textit{A nonsingular Friedmann cosmology}, Sov. Astron. Lett. {\bf 1} (1975) 93
\bibitem{Sato:1980yn}
K.~Sato, \textit{First Order Phase Transition Of A Vacuum And Expansion Of The Universe}, Mon.\ Not.\ Roy.\ Astron.\ Soc.\  {\bf 195} (1981) 467; A.H. Guth,
\textit{The inflationary universe: A possible solution to the horizon and flatness problems},
Phys. Rev. D {\bf 23} (1981) 347;
A. Linde, \textit{A new inflationary universe scenario: A possible solution of the horizon, flatness, homogeneity, isotropy and primordial monopole problems},
Phys. Lett. B {\bf 108} (1982) 389;
A. Albrecht and P.J. Steinhardt,
\textit{Cosmology for grand unified theories with radiatively induced symmetry breaking},
Phys. Rev. Lett. {\bf 48} (1982) 1220.

\bibitem{tony}
D. Lyth and A. Riotto, \textit{Particle Physics Models of Inflation and the Cosmological Density Perturbation}, Phys. Rept. {\bf 314} (1999) 1 [hep-ph/9807278]
\bibitem{f1}G. Dvali, Q. Shafi and R. Schaefer, \textit{Large Scale Structure and Supersymmetric Inflation without Fine Tuning}, Phys. Rev. Lett. {\bf 73} (1994) 1886 [hep-ph/9406319]
\bibitem{f2} E. Copeland, A. Liddle, D. Lyth, E. Stewart and D. Wands, \textit{False Vacuum Inflation with Einstein Gravity}, Phys.Rev. D {\bf 49} (1994) 6410 [astro-ph/9401011]

\bibitem{ks}
 K. Kadota and E. Stewart, \textit{Successful Modular Cosmology}, JHEP {\bf 0307 }(2003) 013 [hep-ph/0304127]; \textit{Inflation on Moduli Space and Cosmic Perturbations}, JHEP {\bf 0312} (2003) 008 [hep-ph/0311240]


\bibitem{hal}
E. Halyo, \textit{Hybrid Inflation from Supergravity D Terms}, Phys. Lett. B {\bf 387} (1996) 43 [hep-ph/9606423]
\bibitem{dvali}
P. Binetruy and G. Dvali, \textit{D-Term Inflation}, Phys. Lett. B {\bf 388 }(1996) 241 [hep-ph/9606342]
\bibitem{casa}
J. Casas, J. Moreno, C. Munoz and M. Quiros, \textit{Cosmological Implications of an Anomalous U(1): Inflation, Cosmic Strings and Constraints on Superstring Parameters}, Nucl. Phys. B {\bf 328 }(1989) 272; J. Casas and C. Munoz, \textit{Inflation from Superstrings}, Phys. Lett. B {\bf 216 }(1989) 37



\bibitem{ewan}
E. Stewart, \textit{Inflation, Supergravity and Superstrings}, Phys. Rev. D {\bf 51 }(1995) 6847 [hep-ph/9405389]

\bibitem{seesaw}
T. Yanagida, \textit{Horizontal Symmetry and Masses of Neutrinos}, Prog. Theor. Phys. {\bf 64 }(1980) 1103; M. Gell-Mann, P. Ramond and R. Slansky, \textit{Complex Spinors and Unified Theories}, published in Supergravity, P. van Nieuwenhuizen and D.Z. Freedman (eds.), North Holland Publ. Co., (1979).




\bibitem{yokoyama}
H. Murayama, H. Suzuki, T. Yanagida and J. Yokoyama, \textit{Chaotic Inflation and Baryogenesis by Right-handed Sneutrinos},  Phys. Rev. Lett. {\bf 70 }(1993) 1912 [hep-ph/9311326]; \textit{Chaotic Inflation and Baryogenesis in Supergravity}, Phys.Rev. D {\bf 50 }(1994) 2356 [hep-ph/9311326]; H. Murayama and T. Yanagida, \textit{Leptogenesis in Supersymmetric Standard Model with Right-handed Neutrino}, Phys. Lett. B {\bf 322 }(1994) 349 [hep-ph/9310297]
\bibitem{ellis}
J. Ellis, M. Raidal and T. Yanagida, \textit{Sneutrino Inflation in the Light of WMAP: Reheating, Leptogenesis and Flavor-Violating Lepton Decays}, Phys. Lett. B {\bf 581 }(2004) 9 [hep-ph/0303242]







\bibitem{lindehy}
 A. Linde, \textit{Hybrid Inflation}, Phys. Rev. D {\bf 49 }(1994) 748 [astro-ph/9307002]
\bibitem{shafi1}
S. Antusch, M. Bastero-Gil, S. King and Q. Shafi, \textit{Sneutrino Hybrid Inflation in Supergravity}, Phys. Rev. D {\bf 71 }(2005) 083519 [hep-ph/0411298]


\bibitem{sy}
O. Seto and J. Yokoyama, \textit{Hiding cosmic strings in supergravity D-term inflation} [hep-ph/0508172]


\bibitem{coleman}
S. Coleman and E. Weinberg, \textit{ Radiative Corrections as the Origin of Spontaneous Symmetry Breaking}, Phys. Rev. D {\bf 7 }(1973) 1888



\bibitem{string}
R. Jeannerot, \textit{Inflation in Supersymmetric Unified Theories}, Phys. Rev. D {\bf 56  }(1997) 6205 [hep-ph/9706391]; D. Lyth and A. Riotto, \textit{Comments on D-term inflation}, Phys. Lett. B {\bf 412 }(1997) 28 [hep-ph/9707273]

\bibitem{rocher}J. Rocher and M. Sakellariadou,\textit{Constraints on Supersymmetric Grand Unified Theories from Cosmology}, JCAP {\bf 0503 }(2005) 004 [hep-ph/0406120]



\bibitem{endo}
 M. Endo, M. Kawasaki and T. Moroi, \textit{Cosmic String from D-term Inflation and Curvaton}, Phys. Lett. B {\bf 569 }(2003) 73 [hep-ph/0304126]



\bibitem{fuku}
M. Fukugita and T. Yanagida, \textit{Baryogenesis without Grand Unification}, Phys. Lett. B {\bf 174 }(1986) 45


\bibitem{cp}
L. Covi, E. Roulet and F. Vissani, \textit{CP violating decays in leptogenesis scenarios}, Phys. Lett. B {\bf 384 }(1996) 169 [hep-ph/9605319];
J. Liu and G. Segre, \textit{Re-Examination of Generation of Baryon and Lepton Number Asymmetries by Heavy Particle Decay}, Phys. Rev. D {\bf 48 }(1993) 4609 [hep-ph/9304241]
\bibitem{olive1}
B. Campbell, S. Davidson and K. Olive, \textit{Inflation, Neutrino Baryogenesis, and (S)Neutrino-Induced Baryogenesis}, Nucl. Phys. B {\bf 399 }(1993) 111 [hep-ph/9302223]


\bibitem{sk}
Super-Kamiokande Collaboration, \textit{A Measurement of Atmospheric Neutrino Oscillation Parameters by Super-Kamiokande I}, Phys. Rev. D {\bf 71 }(2005) 112005 [hep-ex/0501064]


\bibitem{res}
A. Pilaftsis, \textit{Heavy Majorana Neutrinos and Baryogenesis}, Int. J. Mod. Phys. A {\bf 14 }(1999) 1811 [hep-ph/9812256]; 
A. Pilaftsis and T. Underwood, \textit{Resonant Leptogenesis}, Nucl. Phys. B {\bf 692 }(2004) 303 [hep-ph/0309342]





\bibitem{marco}
H. Nilles, M. Peloso and L. Sorbo, \textit{Coupled fields in external background with application to nonthermal production of gravitinos}, JHEP {\bf 0104 }(2001) 004  [hep-th/0103202]; R. Kallosh, L. Kofman, A. Linde and A. Proeyen, \textit{ Gravitino Production after Inflation}, Phys. Rev. D {\bf 61 }(2000) 103503 [hep-th/9907124]; G. Giudice, I. Tkachev and A. Riotto, \textit{ Nonthermal Production of Dangerous Relics in the Early Universe}, JHEP {\bf 9908 }(1999) 009 [hep-ph/9907510]; P. Greene, K. Kadota and H. Murayama, \textit{Supergravity Inflation Free from Harmful Relics}, Phys. Rev. D {\bf 68 }(2003) 043502 [hep-ph/0208276]; 
K.~Kohri, M.~Yamaguchi and J.~Yokoyama, \textit{Production and dilution of gravitinos by modulus decay}, Phys.\ Rev.\ D {\bf 70} (2004) 043522 [hep-ph/0403043]; K.~Kohri, M.~Yamaguchi and J.~Yokoyama, \textit{ Neutralino dark matter from heavy gravitino decay}, Phys.\ Rev.\ D {\bf 72} (2005) 083510  [hep-ph/0502211]



\bibitem{moroi}
M. Kawasaki, K. Kohri and T. Moroi, \textit{Hadronic Decay of Late-Decaying Particles and Big-Bang Nucleosynthesis}, Phys. Lett. B {\bf 625 }(2005) 7 [astro-ph/0402490]; \textit{Big-Bang Nucleosynthesis and Hadronic Decay of Long-Lived Massive Particles}, Phys. Rev. D {\bf 71 }(2005) 083502 [astro-ph/0408426]; K. Jedamzik, \textit{Did Something Decay, Evaporate, or Annihilate during Big Bang Nucleosynthesis?}, Phys. Rev. D {\bf 70 }(2004) 063524 [astro-ph/0402344]

\bibitem{oldgrav1}
S. Weinberg, \textit{Cosmological Constraints on the Scale of Supersymmetry Breaking}, Phys. Rev. Lett. {\bf 48} (1982) 1303; D. Lindley, \textit{Cosmological Constraints On The Lifetime Of Massive Particles}, Astrophys. J. {\bf 294} (1985) 1; M. Khlopov and A. Linde, \textit{Is It Easy To Save The Gravitino?}, Phys. Lett. B {\bf 138} (1984) 265; J. Ellis, J. Kim and D. Nanopoulos, \textit{Cosmological Gravitino Regeneration And Decay}, Phys. Lett. B {\bf 145} (1984) 181; M. Kawasaki and K. Sato, \textit{Decay Of Gravitinos And Photodestruction Of Light Elements}, Phys. Lett. B {\bf 189} (1987) 23; J. Ellis, G. Gelmini, J. Lopez, D. Nanopoulos and S. Sarkar, \textit{Astrophysical Constraints On Massive Unstable Neutral Relic Particles}, Nucl. Phys. B {\bf 373} (1992) 399
   
\bibitem{grav2}
 M. Reno and D. Seckel, \textit{Primordial nucleosynthesis: The effects of injecting hadrons}, Phys. Rev. D {\bf 37} (1988) 3441; S. Dimopoulos, R. Esmailzadeh, L. Hall and G. Starkman, \textit{Is the universe closed by baryons? Nucleosynthesis with a late-decaying massive particle}, Astrophys.J. {\bf 330} (1988) 545


\bibitem{mike}
J. Harvey and M. Turner, \textit{Cosmological Baryon and Lepton Number in the Presence of Electroweak Fermion Number Violation}, Phys. Rev. D {\bf 42 }(1990) 3344



\bibitem{kolda}
C. Kolda and J. March-Russell, \textit{Supersymmetric D-term Inflation, Reheating and Affleck-Dine Baryogenesis}, Phys. Rev. D {\bf 60 }(1999) 023504 [hep-ph/9802358]



\bibitem{ndominated}
K. Hamaguchi, H. Murayama and T. Yanagida, \textit{Leptogenesis from $\widetilde{N}$-dominated early universe}, Phys. Rev. D {\bf 65 }(2002) 043512 [hep-ph/0109030]





















\bibitem{linde1}
J. Garcia-Bellido and A. Linde, \textit{Preheating in Hybrid Inflation}, Phys. Rev. D {\bf 57} (1998) 6075 [hep-ph/9711360]

\bibitem{turb}
R. Micha and I. Tkachev, \textit{Turbulent Thermalization}, Phys. Rev. D {\bf 70 }(2004) 043538 [hep-ph/0403101]













\end{thebibliography}
\end{document}